\documentclass[pra,twocolumn,showpacs]{revtex4}
\usepackage{graphicx}

\newcommand{\beq}{\begin{equation}}
\newcommand{\eeq}{\end{equation}}
\newcommand{\beqa}{\begin{eqnarray}}
\newcommand{\eeqa}{\end{eqnarray}}
\newcommand{\ba}{\begin{array}}
\newcommand{\ea}{\end{array}}

\begin{document}

\title{Bose-Einstein condensates under a spatially-modulated transverse
confinement}
\author{L. Salasnich$^{1,2}$, A. Cetoli$^{2}$, B.A. Malomed$^{3}$, F. Toigo$%
^{1,2}$, and L. Reatto$^{4}$}
\affiliation{$^1$CNISM and CNR-INFM, Unit\`a di Padova, Via Marzolo 8, 35131 Padova,
Italy \\
$^{2}$Dipartimento di Fisica ``Galileo Galilei'', Universit\`a di Padova,
Via Marzolo 8, 35131 Padova, Italy \\
$^{3}$Department of Interdisciplinary Studies, School of Electrical
Engineering, Faculty of Engineering, Tel Aviv University, Tel Aviv 69978,
Israel \\
$^{4}$Dipartimento di Fisica and CNISM, Universit\`a di Milano, Via Celoria
16, 20133 Milano, Italy}

\begin{abstract}
We derive an effective nonpolynomial Schr\"{o}dinger equation (NPSE) for
self-repulsive or attractive BEC in the nearly-1D cigar-shaped trap, with
the transverse confining frequency periodically modulated along the axial
direction. Besides the usual linear cigar-shaped trap, where the periodic
modulation emulates the action of an optical lattice (OL), the model may be
also relevant to toroidal traps, where an ordinary OL cannot be created. For
either sign of the nonlinearity, extended and localized states are found, in
the numerical form (using both the effective NPSE and the full 3D
Gross-Pitaevskii equation) and by means of the variational approximation
(VA). The latter is applied to construct ground-state solitons and predict
the collapse threshold in the case of self-attraction. It is shown that
numerical solutions provided by the one-dimensional NPSE are always very
close to full 3D solutions, and the VA yields quite reasonable results too.
The transition from delocalized states to gap solitons, in the first finite
bandgap of the linear spectrum, is examined in detail, for the repulsive and
attractive nonlinearities alike.
\end{abstract}

\maketitle

\section{Introduction}

It is well known that Bose-Einstein condensates (BECs) with weak attractive
interactions between atoms can form stable solitons in ``cigar-shaped"
(nearly one-dimensional, 1D) traps. In these traps, the gas is strongly
bound in the transverse plane, while being loosely confined along the
longitudinal axis ($z$). Using this configuration, stable bright solitons
were created in the gas of $^{7}$Li atoms \cite{Strecker02,Khaykovich02}. In
the condensate of $^{85}$Rb atoms trapped in a similar configuration,
stronger attraction between atoms leads to collapse and emergence of nearly
3D solitons in a \textit{post-collapse} state \cite{Cornish}.

The strongly elongated (cigar-shaped) settings are described by effectively
1D equations which were derived, by dint of various approximations, from the
full 3D Gross-Pitaevskii equation (GPE) \cite{PerezGarcia98}-\cite{denicola}%
. The derivation assumes an \textit{ansatz} factorizing the 3D wave function
into a product of a transverse mode and an axial (one-dimensional) wave
function. As shown in Refs. \cite{sala1,sala2}, the substitution of the
factorized ansatz in the underlying cubic GPE leads, in the general
situation, to a \textit{nonpolynomial Schr\"{o}dinger equation} (NPSE) for
the axial wave function (or a system of coupled NPSEs for a two-component
BEC \cite{we-first}). In the case of weak nonlinearity, the NPSE can be
expanded, which leads to a simplified 1D equation with a combination of
cubic and quintic terms, the latter one being always \textit{attractive}; if
the cubic term is attractive too, the cubic-quintic equation gives rise to a
family of stable solitons available in an exact analytical form \cite{Lev}.
Unlike the ordinary 1D nonlinear Schr\"{o}dinger (NLS)\ equation with the
attractive cubic term, the NPSE, as well as its cubic-quintic truncation,
may give rise to collapse, which reflects the occurrence of the collapse in
the underlying cubic GPE in three dimensions \cite{sala1,sala2,Lev}. Despite
the possibility of the collapse, solitons are stable in these models.

A relevant problem is to derive the NPSE for the cigar-shaped trap equipped
with a periodic potential, which is created in the experiment as an \textit{%
optical lattice} (OL), i.e., an interferences pattern, by a pair of
counterpropagating laser beams illuminating the trap in the axial direction
(the BEC dynamics in periodic potentials was recently reviewed in Ref. \cite%
{Markus}). A 1D equation of the NPSE type including the OL potential was
recently derived in Ref. \cite{we}. Using that equation, the influence of
the periodic potential on the collapse threshold for axially localized
states in the quasi-1D trap was investigated, both for single-peak solitons
found in the semi-infinite gap of the periodic potential, and multi-peaked
solitons found in finite bandgaps. The results were compared to direct
numerical solutions of the full 3D GPE, showing good agreement.

In the experiment, the transverse potential which confines the atomic gas to
the cigar-shaped configuration may be axially nonuniform, which corresponds
to the corresponding trapping frequency being a function of the axial
coordinate, $\Omega _{\perp }=\Omega _{\perp }(z)$ (generally speaking, it
may also depend on time). As proposed in Ref. \cite{denicola}, a specially
designed nonuniformity (\textit{axial modulation}) of $\Omega _{\perp }$ may
be used as an alternative tool for the control of dynamics of nearly 1D
solitons, inducing an effective potential for them. In that work, the
effective potential for a soliton with norm (scaled number of atoms) $N$ was
found in the limit of weak nonlinearity and long-scale axial modulation,
\begin{equation}
U_{\mathrm{eff}}^{\mathrm{(sol)}}(\zeta )=\left( 1+S\right) N\Omega _{\perp
}(\zeta )+\mathcal{O}\left( \gamma ^{2}N^{3}\right) ,  \label{eff}
\end{equation}%
where $\gamma $ is an effective nonlinearity coefficient, see Eq. (\ref{efun}%
) below (we will use normalization tantamount to setting $N\equiv 1$), $%
\zeta $ is the coordinate of the soliton's center, and integer $S$ is
possible \textit{intrinsic vorticity} of the quasi-1D soliton, which is
defined below in Eq. (\ref{S}) (ordinary solitons correspond to $S=0$).
Expression (\ref{eff}) was derived by means of a variational approximation
(VA) applied directly to the full 3D energy functional, cf. Eq. (\ref{e-npse}%
) below.

Of special interest is the case of periodic modulation of the transverse
trapping frequency, with wavenumber $k$, modulation depth $\alpha <1$, and
amplitude $\omega _{\perp }$:
\begin{equation}
\Omega _{\perp }^{2}(z)=\omega _{\bot }^{2}\left[ 1-\alpha \cos {(2kz)}%
\right] .  \label{U}
\end{equation}%
As suggested by Eq. (\ref{eff}), the periodic modulation may replace the OL
potential. In particular, this setting may be especially relevant to a
situation when the quasi-1D magnetic trap is not rectilinear, but rather
circular (toroidal), which was realized in the experiment \cite{torus}.
Indeed, the OL cannot be created in such a setting, but the induction of an
effective periodic potential by means of the modulation of the transverse
trapping frequency is quite feasible.

The objective of the present work is to derive an effective NPSE for the
quasi-1D trap subject to the periodic modulation as per Eq. (\ref{U}), and
to investigate various self-trapped states in this geometry, both
delocalized ones and solitons. The paper is organized as follows. The NPSE
is derived, in a general form, in Sec. II. Then, the model with the
repulsive nonlinearity is considered in Sec. III. At first, the analysis
includes an additional parabolic trapping potential acting in the axial
direction. Then, this potential is removed, and we analyze solutions
demonstrating a transition from delocalized solutions to a gap soliton. The
solutions predicted by the effective 1D equation are compared to their
counterparts found from the full 3D GPE (bandgap structures generated by
linearized versions of both equations are compared too). Section IV\ is
dealing with the case of the attractive nonlinearity. The corresponding
ground-state solitons are found by means of the VA and in a numerical form,
which are also compared with results produced by the full 3D equations.
Because the attractive nonlinearity may give rise to collapse, the collapse
threshold is considered in detail too, by means of both the VA and numerical
methods. In addition to the ground-state solitons, which reside in the
semi-infinite gap induced by periodic modulation (\ref{U}), we also explore
solutions with the chemical potential belonging to the first finite gap, and
demonstrate the transition from delocalized states to gap solitons in that
case. The paper is concluded by Sec. V.

\section{The nonpolynomial Schr\"{o}dinger equation}

Static BEC configurations can be derived from the normalized functional
which determines the energy-per-atom in terms of order parameter
(single-atom wave function) $\psi (\mathbf{r})$ and includes a generic
external axial potential $V(z)$:
\[
E=\int d^{3}\mathbf{r}\,\psi ^{\ast }(\mathbf{r})\left[ -{\frac{1}{2}}\nabla
^{2}+{\frac{1}{2}}\left[ 1-\alpha \cos {(2kz)}\right] (x^{2}+y^{2})\right.
\]%
\begin{equation}
\left. +V(z)+\pi \gamma |\psi (\mathbf{r})|^{2}\right] \psi (\mathbf{r}).
\label{efun}
\end{equation}%
Here $\gamma \equiv 2a_{s}N/a_{\bot }$ is the adimensional strength of the
interaction between atoms, with $a_{s}$ the inter-atomic scattering length
and $N$ the number of atoms in the condensate. In Eq. (\ref{efun}) lengths
are measured in units of the characteristic transverse trapping
(harmonic-oscillator) length, $a_{\bot }=\sqrt{\hbar /(m\omega _{\bot })}$,
where $m$ is the atomic mass, the energy is in units of $\hbar \omega _{\bot
}$, modulation wave number $k$ is in units of $a_{\bot }^{-1}$, and the wave
function is subject to the ordinary normalization,
\begin{equation}
\int |\psi (\mathbf{r})|^{2}d\mathbf{r}\equiv 1.  \label{N}
\end{equation}%
The chemical potential corresponding to Eqs. (\ref{efun}) and (\ref{N}) is
\begin{equation}
\mu =E+\pi \gamma \int \left\vert \psi (\mathbf{r})\right\vert ^{4}d\mathbf{%
r.}  \label{mu}
\end{equation}%
Due to the above normalizations, $\omega _{\perp }$ and $N$ are not
explicitly present in expressions (\ref{efun}) and (\ref{mu}).

An accurate investigation of the present setting can be performed by using
the approach which was first developed for the GPE with the unmodulated
transverse trap; after averaging the full 3D equation in the transverse
plane, it leads to an effective one-dimensional NPSE \cite{sala1}. The
derivation of the NPSE starts with the factorization of the 3D wave function
into a product of an arbitrary complex axial wave function, $f(z)$, and the
ordinary transverse Gaussian \textit{ansatz} with transverse width $\sigma
(z)$,
\begin{equation}
\psi (\mathbf{r})={\frac{1}{\sqrt{\pi }\sigma (z)}}\exp {\left\{ -{\frac{%
(x^{2}+y^{2})}{2\sigma ^{2}(z)}}\right\} }\,f(z),  \label{factor}
\end{equation}%
$f(z)$ being subject to normalization condition
\begin{equation}
\int_{-\infty }^{+\infty }dz\,|f(z)|^{2}=1.  \label{N1D}
\end{equation}

Configurations including the above-mentioned intrinsic vorticity,
characterized by positive integer $S$, correspond to the following
generalization of Eq. (\ref{factor}):%
\begin{equation}
\psi (\mathbf{r})=\frac{1}{\sqrt{\pi S!}\sigma^{1+S}(z)}\rho^{S}\exp {%
\left\{ -{\frac{\rho ^{2}}{2\sigma ^{2}(z)}+iS\theta }\right\} }\,f(z),
\label{S}
\end{equation}%
where $\rho \equiv \sqrt{x^{2}+y^{2}}$ and $\theta $ are the polar
coordinates in transverse plane $\left( x,y\right) $. Nearly-1D solitons
with intrinsic vorticity were studied, by means of methods similar to those
used in the present work (although without deriving a closed-form NPSE for
that case) in Ref. \cite{Luca}; for a limit case of a strongly elongated 3D
trap, similar localized states were also studied in Ref. \cite{Dumitru}. In
this work, we focus on the ordinary solitons, with $S=0$.

Inserting expression (\ref{factor}) in Eq. (\ref{efun}), one obtains,
neglecting the $z$-derivative of $\sigma$, 
the following effective (1D) energy functional: 
\[
E=\int dz\,f^{\ast }(z)\left\{ -{\frac{1}{2}}{\frac{\partial ^{2}}{\partial
z^{2}}}+V(z)+{\frac{1}{2}}\left( {\frac{1}{\sigma ^{2}(z)}}+\right. \right.
\]%
\begin{equation}
\left. \left. \left[ 1-\alpha \cos {(2kz)}\right] \sigma ^{2}(z)\right) +{%
\frac{1}{2}}{\frac{\gamma }{\sigma ^{2}(z)}}|f(z)|^{2}\right\} f(z)~.
\label{e-npse}
\end{equation}%
The minimization of this functional with respect to $f(z)$, taking
normalization (\ref{N1D}) into regard, leads to the stationary NPSE,
\[
\left[ -{\frac{1}{2}}{\frac{d^{2}}{dz^{2}}}+V(z)+{\frac{1}{2}}\left( {\frac{1%
}{\sigma ^{2}(z)}}+\left[ 1-\alpha \cos {(2kz)}\right] \sigma ^{2}(z)\right)
\right.
\]%
\begin{equation}
\left. +{\frac{\gamma }{\sigma ^{2}(z)}}|f(z)|^{2}\right] f=\mu \,f.
\label{npse}
\end{equation}%
where the chemical potential $\mu $ appears as the Lagrange multiplier
generated by the normalization condition.

An equation for the transverse width is obtained by the minimization of
functional (\ref{e-npse}) with respect to $\sigma (z)$:%
\begin{equation}
\sigma ^{2}(z)=\sqrt{{\frac{1+\gamma |f(z)|^{2}}{1-\alpha \cos {(2kz)}}}}.
\label{sigma-npse}
\end{equation}%
The substitution of this expression in Eq. (\ref{npse}) leads to a
closed-form stationary NPSE for $f(z)$, although in a rather cumbersome form.

In the limit of zero scattering length, $\gamma =0$, Eqs. (\ref{npse}) and (%
\ref{sigma-npse}) reduce to the linear Schr\"{o}dinger equation,
\begin{equation}
\left[ -{\frac{1}{2}}{\frac{\partial ^{2}}{\partial z^{2}}}+V(z)+\sqrt{%
1-\alpha \cos {(2kz)}}\right] f(z)=\mu f(z),  \label{npse-g0}
\end{equation}%
with the effective axial potential,
\begin{equation}
V_{\mathrm{eff}}(z)=V(z)+\sqrt{1-\alpha \cos {(2kz)}}  \label{Veff}
\end{equation}%
(note that, even in this limit, the full 3D GPE is not separable, due to the
modulation imposed on the transverse trapping potential). If the nonlinear
term $\gamma |f(z)|^{2}$ is small, the NPSE may be approximated by the cubic
NLS equation with the same effective axial potential and, in addition to
that, with a periodically-modulated nonlinearity coefficient:
\[
\left[ -{\frac{1}{2}}{\frac{\partial ^{2}}{\partial z^{2}}}+V_{\mathrm{eff}%
}(z)\right.
\]%
\begin{equation}
\left. +\gamma \,\sqrt{1-\alpha \cos {(2kz)}}\,|f(z)|^{2}\right] f(z)=\mu
f(z).  \label{NLS}
\end{equation}%
Essentially the same result as given by Eq. (\ref{NLS}) was obtained, in the
simplest approximation, in Ref. \cite{denicola}, see Eqs. (\ref{eff}) and (%
\ref{U}).

\section{The model with repulsive nonlinearity}

\subsection{One-dimensional solutions}

In the case of the repulsive inter-atomic interaction, i.e., positive $a_{s}$
and $\gamma $, the application of the Thomas-Fermi (TF) approximation, which
neglects the second derivative, to Eqs. (\ref{npse}) and (\ref{sigma-npse})
yields an analytical expression for the normalized atomic density, $\rho
(z)\equiv |f(z)|^{2}$:
\begin{equation}
\rho (z)={\frac{2}{9\gamma }}\left[ \mu _{\mathrm{eff}}^{2}(z)-3+\mu _{%
\mathrm{eff}}(z)\sqrt{\mu _{\mathrm{eff}}^{2}(z)+3}\right] ,  \label{figata}
\end{equation}%
where an effective local chemical potential is
\begin{equation}
\mu _{\mathrm{eff}}(z)={\frac{\mu -V(z)}{\sqrt{1-\alpha \cos {(2kz)}}}.}
\label{figata1}
\end{equation}%
Equations (\ref{figata}) and (\ref{figata1}) generalize the result obtained
in Ref. \cite{sala3} for the repulsive BEC under the unmodulated transverse
confinement ($\alpha =0$). In the limit case of strong nonlinearity, $\gamma
\rho \gg 1$, Eq. (\ref{figata}) reduces to $\rho (z)=\left( 4/9\gamma
\right) \mu _{\mathrm{eff}}^{2}(z),$and in the opposite limit of $\gamma
\rho \ll 1$, which implies $\mu _{\mathrm{eff}}(z)-1\ll 1$, it takes the
form $\rho (z)=\left( 1/\gamma \right) (\mu _{\mathrm{eff}}(z)-1)$.

To obtain accurate results, we solved the time-dependent variety of Eq. (\ref%
{npse}) (with $\mu $ replaced by $i\partial /\partial t$ and, accordingly, $%
d/dz$ replaced by $\partial /\partial z$), combined with Eq. (\ref%
{sigma-npse}) (without any change in the latter equation) numerically, by
means of the finite-difference Crank-Nicholson method in imaginary time,
following the approach elaborated in Ref. \cite{sala-numerics}. The explicit
axial potential was chosen in the usual form, $V(z)=\left( \lambda
\,z\right) ^{2}/2$, where $\lambda \equiv \omega _{z}/\omega _{\bot }$, and $%
\omega _{z}$ is the axial-confinement frequency. In this way, the profiles
for $\rho (z)$, displayed by dashed lines in Fig. 1, have been obtained for
different values of modulation depth $\alpha $ and fixed nonlinearity
strength, $\gamma =20$. The so obtained NPSE profiles are compared to those
produced by the TF approximation (dotted lines in Fig. 1), see Eq. (\ref%
{figata}). In the absence of transverse modulation, $\alpha =0$ (the upper
panel in Fig. 1), the axial density profile is well approximated by the TF
formula. At $\alpha \neq 0$ (the central and lower panels in Fig. 1), the TF
approximation much overestimates the density contrast between points of
local minima and maxima of the total axial potential.

\begin{figure}[tbp]
{\includegraphics[height=3.0in,clip]{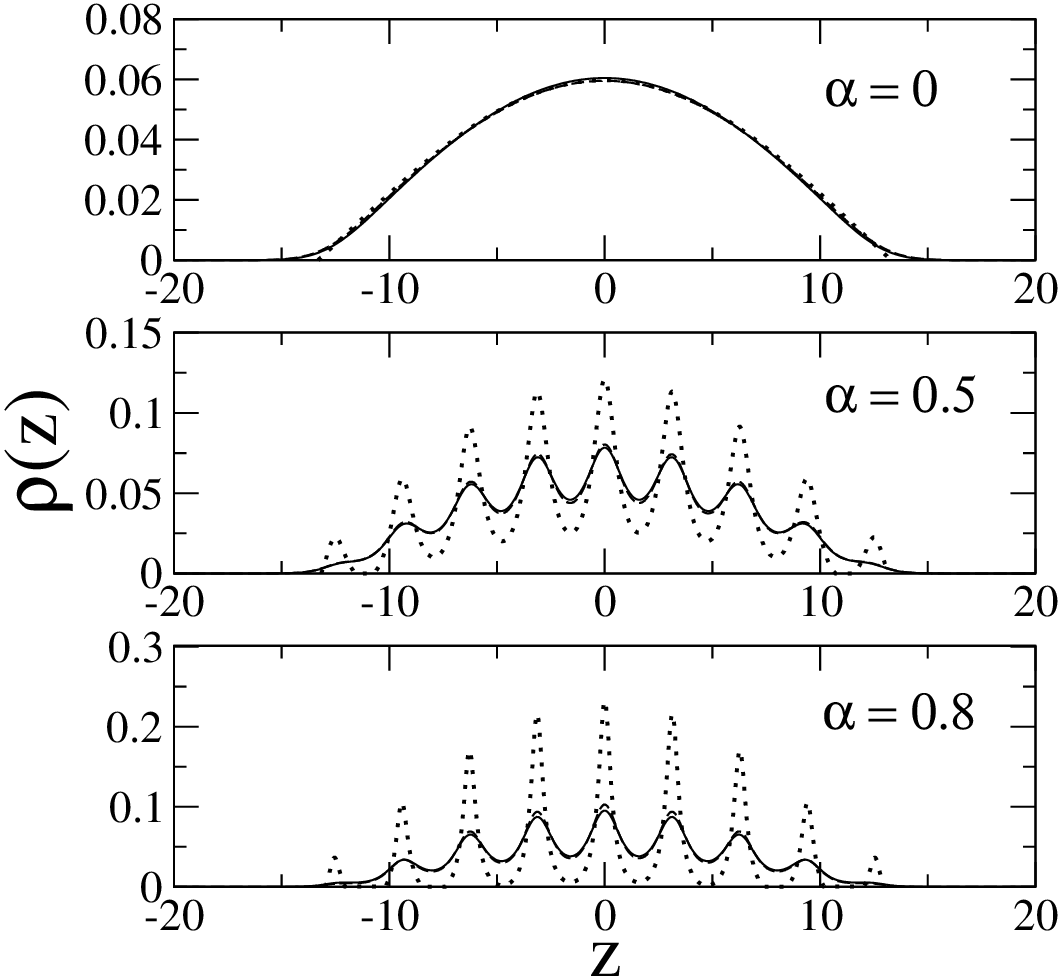}}
\caption{The axial profile of density $\protect\rho (z)$ in the
self-repulsive BEC with $\protect\gamma =20$, $\protect\lambda =0.1$, $k=1$,
at different values of the modulation amplitude $\protect\alpha $. Solid
lines: results obtained from the full three-dimensional GPE, derived by the
minimization of the underlying energy functional, which is given by Eq. (%
\protect\ref{efun}). Dashed lines: the profiles produced by the numerical
solution of the one-dimensional \textit{nonpolynomial Schr\"{o}dinger
equation} (NPSE), Eq. (\protect\ref{npse}). Dotted lines: results obtained
from the Thomas-Fermi approximation applied to the NPSE, i.e. Eq. (\protect
\ref{figata}). When using the 3D equation, in this figure and below, the
axial density displayed in the plots is defined by integration in the
transverse plane, $\protect\rho (z)=\protect\int \protect\int |\protect\psi (%
\mathbf{r})|^{2}dxdy$, while in other cases it is simply $|f(z)|^{2}$.}
\end{figure}

\subsection{Three-dimensional solutions}

It is necessary to compare results produced by the NPSE to their
counterparts found from a direct numerical solution of the full GPE in three
dimensions. The latter equation is obtained by the minimization of energy
functional (\ref{efun}). Figure 1 shows that the density profiles generated
by the NPSE (dashed lines) are very close to ones obtained from the 3D
equation (solid lines), unless $\alpha $ is very large. It is noteworthy
that the NPSE gives very accurate results for a model with non-separable
potential.

One may expect that the effective axial periodic potential induced by the
transverse modulation (without the inclusion of the axial parabolic trap)
may support quasi-periodic Bloch states and \emph{gap solitons} in the
self-repulsive BEC, as in the case of the ordinary (direct) periodic
potential \cite{GSprediction,Markus}. To consider this possibility, a
self-consistent numerical method was used, with periodic boundary
conditions. We employed a spatial grid of $1025$ points, covering the
interval of $-50.26\leq z\leq 50.26$, which corresponds to $32$ periods of
the external modulation. To test the numerical scheme, we have checked that
the lowest-energy state in the semi-infinite bandgap (i.e., the ground state
of the system), produced by this method, is identical to that found above by
the integration of the time-dependent NPSE in imaginary time.

In the upper panel of Fig. 2 we plot, as a function of nonlinearity strength
$\gamma $, the first $50$ eigenvalues $\mu _{j}$, as found by means of the
above-mentioned numerical method from Eqs. (\ref{npse}) and (\ref{sigma-npse}%
) with $V(z)=0$ and periodic boundary conditions. The lowest $32$
eigenstates $f_{j}(z)$ belong to the first \textit{nonlinear} band induced
by the periodic modulation in the full NPSE, and the other states, which are
well separated by a \textit{bandgap}, form a second nonlinear band. For some
of the intraband states, the numerical method does not converge to a single
configuration, but rather oscillates between two or three configurations
with very close eigenvalues. It is also possible that the nonlinear model
may give rise to intraband states which have no counterparts in the linear
limit. This challenging issue needs special treatment, and will be
considered elsewhere. As concerns the nonlinear eigenstates for which our
presently employed numerical method leads to convergence (including the gap
soliton, see below), we have verified, by direct simulations of the
time-dependent NPSE in real time, that they \emph{all} are stable.

\begin{figure}[tbp]
{\includegraphics[height=2.8in,clip]{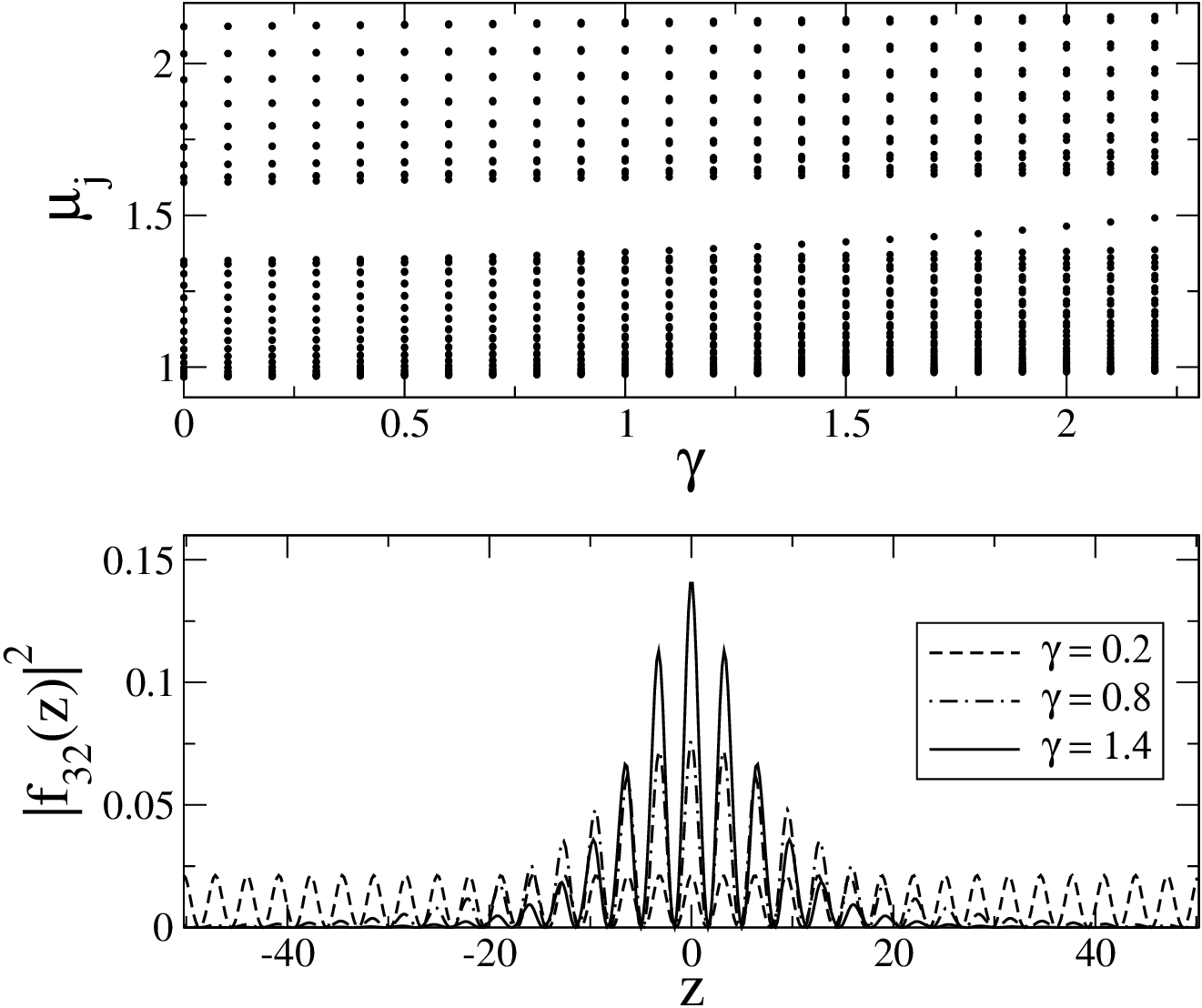}}
\caption{Upper panel: first $50$ eigenvalues $\protect\mu _{j}$ found from
the NPSE, Eq. (\protect\ref{npse}), with $V(z)=0$, $\protect\alpha =0.5$ and
$k=1$, as functions of nonlinearity parameter $\protect\gamma $. Lower
panel: the axial density profile, $|f_{32}(z)|^{2}$, of the 32th
eigenfunction, at $\protect\alpha =0.5$, for three values of $\protect\gamma
$. The figure shows that the 32th eigenstate becomes a \textit{gap soliton}
at sufficiently large $\protect\gamma $.}
\end{figure}

As a typical example, in the lower panel of Fig. 2 we plot the density
profile of the 32th state (it has number $32$ in the set of $50$ numerically
found states) for $\alpha =0.5$ and three different values of $\gamma $. One
can check that, in the linear approximation, this state lies at the top of
the first Bloch band. With the increase of the nonlinearity strength $\gamma$%
, the energy of the 32th state grows, and, in doing so, it enters the first
finite bandgap (as defined in the linear approximation) from below. Figure 2
demonstrates that, for $\gamma =0.2$, this state is still fully delocalized,
being thus similar to a Bloch wave, while for $\gamma =0.8$ it becomes
localized, with a width much smaller than the length of the periodic box.
Clearly, this solution may be identified as a \textit{gap soliton}. At $%
\gamma =1.4$, the gap soliton compresses itself (see Fig. 2) into a still
narrower state.

\begin{figure}[tbp]
\includegraphics[height=2.6in,clip]{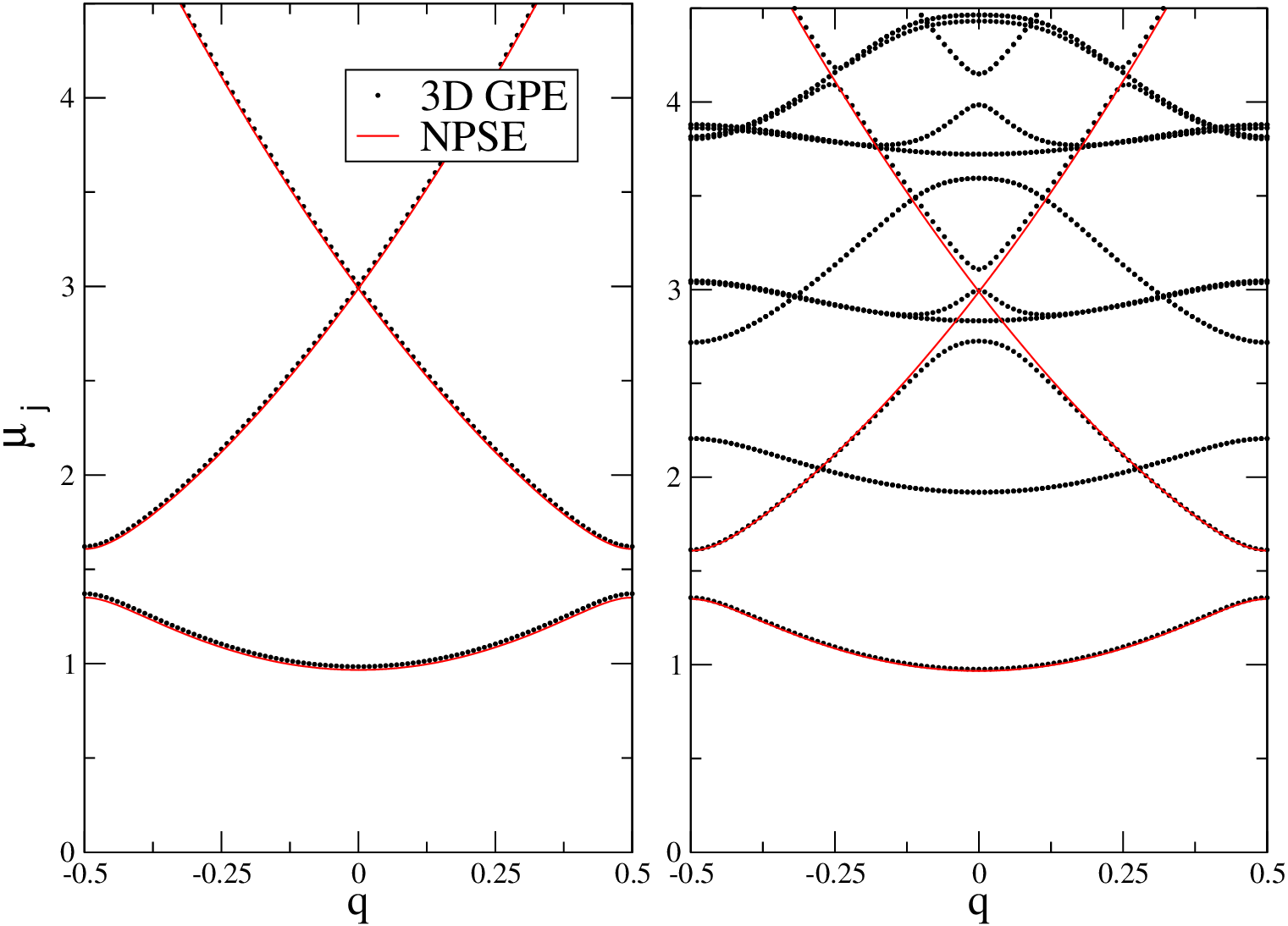}
\caption{Comparison between the dispersion laws as obtained from the full 3D
equation, Eq. (\protect\ref{alce_eigenprob}), and from the effective 1D
equation, Eq. (\protect\ref{npse}), for $k=1$, $\protect\alpha =0.5$ and $%
\protect\gamma =0$. Left panel: Eq. (\protect\ref{alce_eigenprob}) was
solved by confining the summation to $m_{x}=m_{y}=0$. Right panel: the
summation range was extended to $0\leq m_{x},m_{y}\leq 10$.}
\label{alce_figure}
\end{figure}

It is also necessary to check that the bandgap structure in the NPSE with
the periodically modulated trapping potential in the linear limit ($\gamma =0
$) is not, by itself, an artifact of the approximation (reduction of the 3D
equation to the 1D limit), but a true feature of the full 3D model. To this
end, we solved the eigenvalue problem for the full linear GPE in three
dimensions,
\[
\left[ \left( -\frac{1}{2}\nabla _{\perp }^{2}+\frac{1}{2}\frac{\partial ^{2}%
}{\partial z^{2}}\right) +\frac{1}{2}\rho ^{2}\right.
\]%
\begin{equation}
-\left. \frac{1}{2}\alpha \,\rho ^{2}\,\cos (2kz)\right] \,\psi (\rho
,z)=\mu _{j}\,\psi (\rho ,z),  \label{alce_gpe3d}
\end{equation}%
where $\nabla _{\perp }$ acts on $x$ and $y$, and, as above, $\rho
^{2}\equiv x^{2}+y^{2}$. The transverse part of solutions to Eq. (\ref%
{alce_gpe3d}) may be taken as an eigenstate of the corresponding 2D harmonic
oscillator, with its quantum numbers $m_{x}$ and $m_{y}$. Defining $%
G_{1}=2\,k$ as the smallest reciprocal lattice vector and writing $%
G_{n}=n\,G_{1}$ ($n$ is an integer), we find that the solution can be
written as
\begin{equation}
\psi (\rho
,z)=\sum_{m_{x},m_{y},G_{n}}C_{m_{x},m_{y},G_{n}}\,F_{m_{x},m_{y}}(\rho
)\,e^{i\,(q+G_{n})\,z},  \label{alce_bloch_cwf}
\end{equation}%
with coefficients $C_{m_{x},m_{y},G_{n}}$ satisfying a linear eigenvalue
problem,
\[
\left[ \left( m_{x}+m_{y}+1\right) +\frac{1}{2}\left( q+G_{n}\right) ^{2}%
\right] \,C_{m_{x},m_{y},q+G_{n}}
\]%
\begin{equation}
-\sum_{m_{x}^{\prime },m_{y}^{\prime },G^{\prime
}}A_{m_{x},m_{y},G_{n}}^{m_{x}^{\prime },m_{y}^{\prime },G^{\prime
}}\,C_{m_{x}^{\prime },m_{y}^{\prime },q+G^{\prime }}=\mu
_{j}\,C_{m_{x},m_{y},q+G_{n}}~,  \label{alce_eigenprob}
\end{equation}%
where $q$ is a wavenumber in the first Brillouin zone, and the matrix $A$ is
\[
A_{m_{x},m_{y},G_{n}}^{m_{x}^{\prime },m_{y}^{\prime },G^{\prime }}=\frac{1}{%
4}\,\alpha \,{(\delta _{G^{\prime },G_{n+1}}+\delta _{G^{\prime },G_{n-1}})}%
\times \left\{ \delta _{m_{x},m_{x}^{\prime }}\right.
\]%
\[
\left[ \delta _{m_{y},m_{y}^{\prime }}\,\left( \frac{1}{2}+m_{y}\right) +%
\frac{1}{2}\left( \sqrt{(m_{y}^{\prime }+1)(m_{y}^{\prime }+2)}\,\delta
_{m_{y},m_{y}^{\prime }+2}\right. \right.
\]%
\begin{equation}
\left. \left. +\sqrt{(m_{y}+1)(m_{y}+2)}\,\delta _{m_{y},m_{y}^{\prime
}-2}\right) \right] +\delta _{m_{y},m_{y}^{\prime }}  \label{alce_A}
\end{equation}%
\[
\left[ \delta _{m_{x},m_{x}^{\prime }}\,\left( \frac{1}{2}+m_{x}\right) +%
\frac{1}{2}\left( \sqrt{(m_{x}^{\prime }+1)(m_{x}^{\prime }+2)}\,\delta
_{m_{x},m_{x}^{\prime }+2}\right. \right.
\]%
\[
\left. \left. \left. +\sqrt{(m_{x}+1)(m_{x}+2)}\,\delta
_{m_{x},m_{x}^{\prime }-2}\right) \right] \right\} .
\]%
We solved Eq. (\ref{alce_eigenprob}) by truncating the sum in expression (%
\ref{alce_bloch_cwf}) to $-5\leq G_{n}\leq 5$. At first, we also truncated
the summation to $m_{x}=m_{y}=0$ (i.e., only the contribution from the
ground state of the transverse potential was taken into account), which
yielded the result shown in the left panel of Fig. \ref{alce_figure}. Then
we found more accurate solutions, by extending the truncated summation to $%
0\leq m_{x},m_{y}\leq 10$ (i.e., including the contribution from excited
transverse states). In that case, Eq. (\ref{alce_eigenprob}) produces the
dispersion relation shown in the right panel of Fig. \ref{alce_figure}.

These results were compared to the dispersion law as found directly from the
NPSE, Eq. (\ref{npse}). We observe that the two dispersion laws are
identical if the 3D analysis is confined to $m_{x}=m_{y}=0$, while they are
different if the effect of states with $m_{x},m_{y}>0$ are considered.
Nevertheless, Fig. \ref{alce_figure} shows that the first band, the first
gap, and half of the second band do not alter essentially, if the NPSE is
replaced by the full 3D equation, which justifies the use of the NPSE
approximation for values of the chemical potential up to the first half of
the second band.

\section{The model with the attractive nonlinearity}

\subsection{Numerical results}

The model with the attractive inter-atomic interactions, i.e. $\gamma <0$,
may be expected to generate bright solitons. We analyzed this possibility
through the numerical solution of the NPSE equation, in the absence of the
direct axial potential, $V(z)=0$. In Fig. 4, we plot axial density profiles
of the thus found solitons for different values of modulation depth $\alpha $
and fixed nonlinearity strength, $g\equiv -\gamma $.

For $\alpha =0$, the soliton's profile, with a single maximum, may be well
fitted by $f(z)=\left( \sqrt{g}/2\right) \mathrm{sech}\left( gz{/2}\right) $%
, which is an asymptotically exact solution (the usual NLS soliton) in the
above-mentioned weakly nonlinear limit corresponding to $gf^{2}\ll 1$,
provided that $z$ varies on the entire real axis. On the other hand, if $z$
belongs to a finite interval, $-L/2<z<L/2$, with periodic boundary
conditions, $f(z+L)=f(z)$, it is known \cite{sala4} that the NPSE with $%
\alpha =0$ yields a spatially uniform ground state, $f(z)\equiv 1/\sqrt{L}$,
for sufficiently weak nonlinearity, $0\leq g<\pi ^{2}/L$; the ground state
develops a spatial structure at $g>\pi ^{2}/L$ . As shown in Fig. 4, for
nonzero $\alpha $ the soliton profile features several local maxima and
minima due to the action of the effective periodic potential. Thus, under
such conditions, the Bose condensate self-traps into a multi-peaked soliton,
which occupies several cells of the periodic modulation.

\begin{figure}[tbp]
{\includegraphics[height=2.6in,clip]{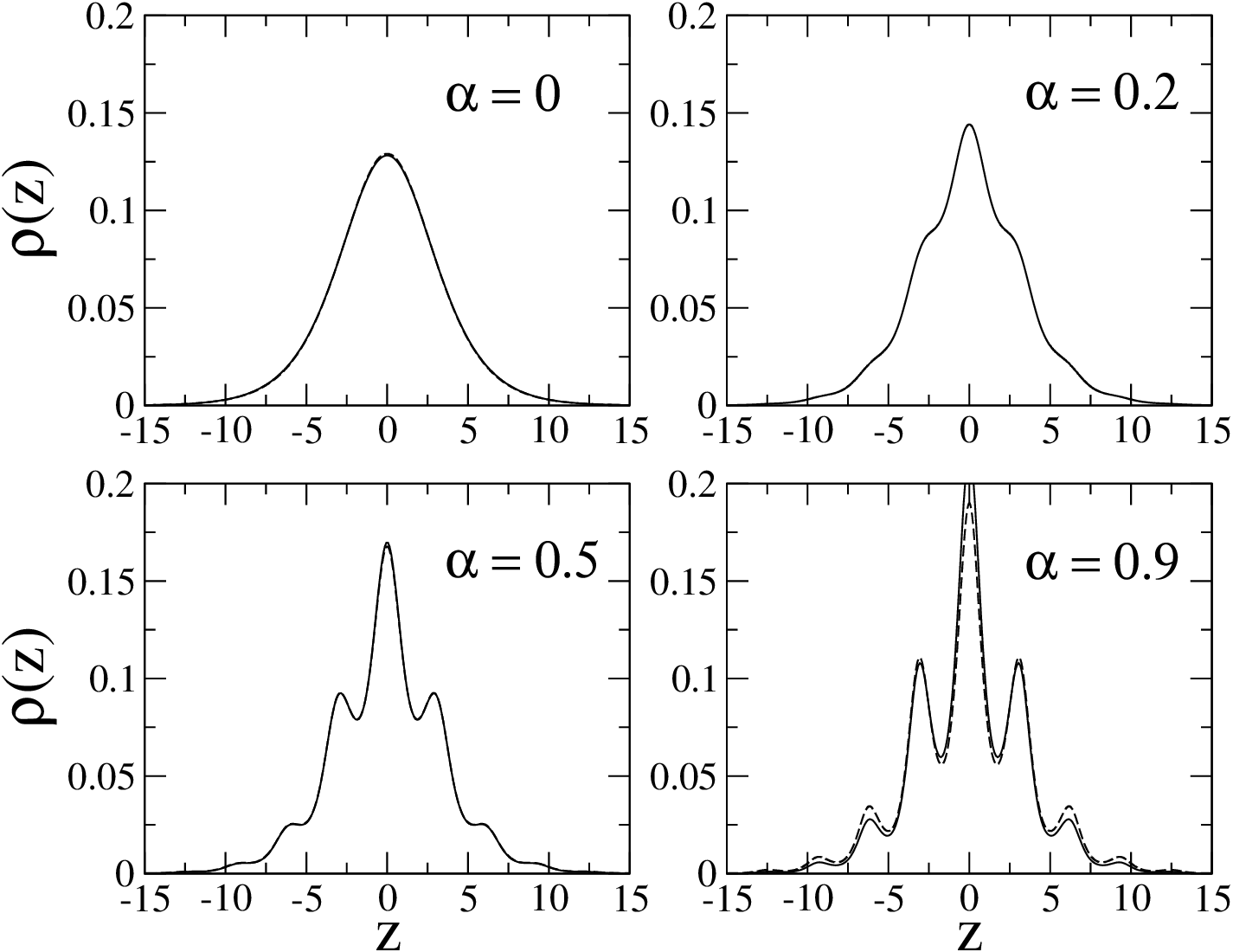}}
\caption{Axial density profiles, $\protect\rho (z)$, of the ground-state
bright soliton in the attractive model with $g=0.5$ and $k=1$, for different
values of modulation amplitude $\protect\alpha $. Dashed lines: results
obtained from the NPSE, Eq. (\protect\ref{npse}). Solid lines: results
obtained from the 3D equation, derived by the minimization of energy
functional (\protect\ref{efun}).}
\end{figure}

\begin{figure}[tbp]
{\includegraphics[height=2.6in,clip]{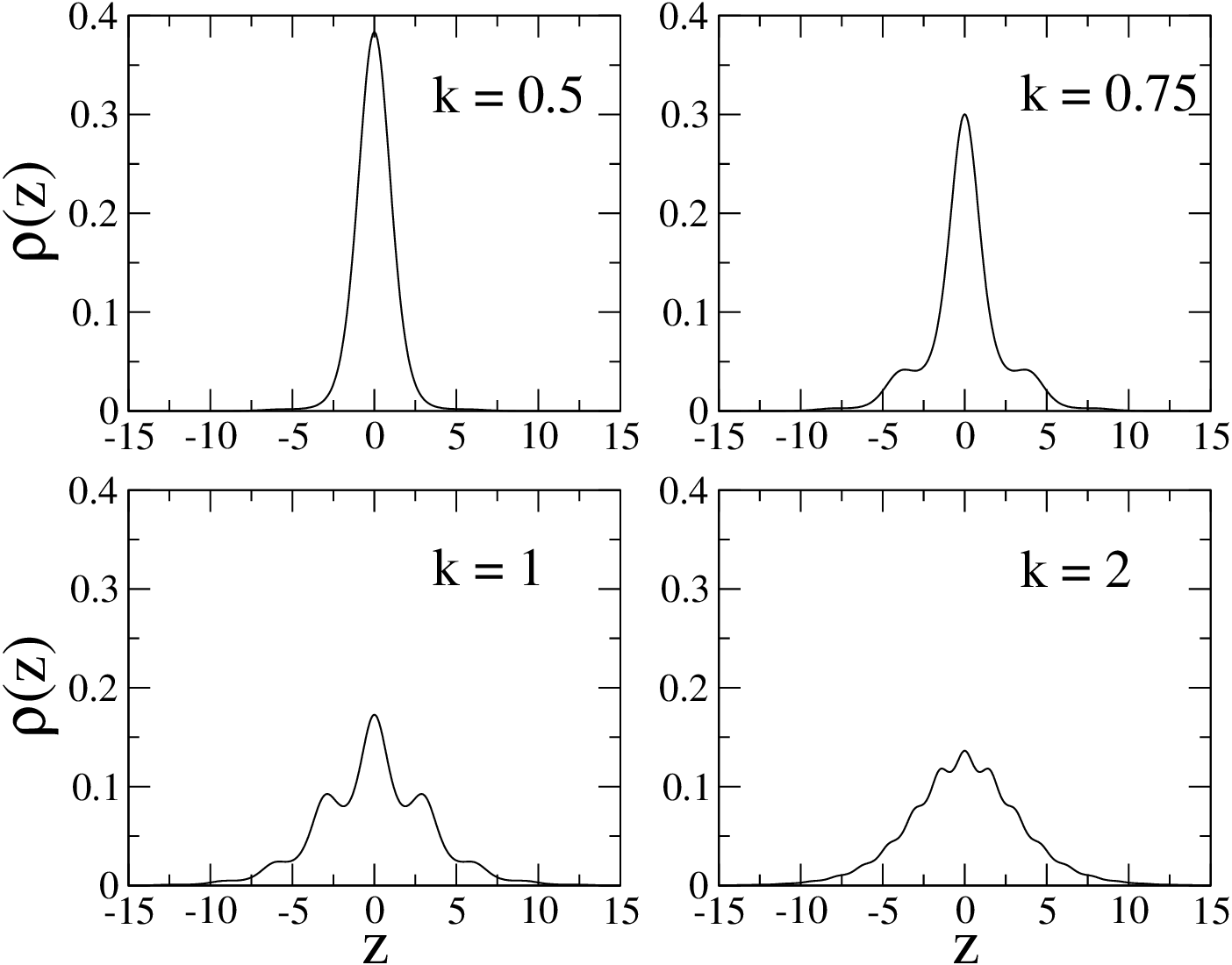}}
\caption{Axial density profiles, $\protect\rho (z)$, of the ground-state
bright soliton in the attractive model with $g=0.5$ and $\protect\alpha =0.5$%
, for different values of wave number $k$. Results obtained from the NPSE,
Eq. (\protect\ref{npse}).}
\end{figure}

We also compared the results yielded by the NPSE with those found from the
direct numerical solution of the full GPE in three dimensions. Figure 4
shows that the density profiles generated by the NPSE (dashed lines)
practically coincide with the ones obtained from the 3D GPE (solid lines),
unless $\alpha $ is very large.

It is also interesting to analyze the behavior of the density profile of the
bright soliton as a function of the wavenumber $k$. In Fig. 5 we plot the
density profile $\rho (z)$ at $g=0.5$ and $\alpha =0.5$, for four values of $%
k$. The figure shows that, at $k=0.5$, the profile is strongly localized
within one single site of the effective periodic axial potential: accordingly, we call the corresponding solution a \textit{single-site soliton%
} (cf. Ref. \cite{we}, where the NPSE for the model with attraction and an
explicit periodic axial potential was considered). At $k \leq 0.75$%
, delocalization of the bright soliton, which occupies more than one site,
is observed. We
call it a \textit{multi-site soliton} (the distinction between the strongly
and weakly localized solitons is also observed in the 1D\ cubic NLS equation
with the self-attractive nonlinearity \cite{NLS}). 

\subsection{Variational vs numerical results}

From the numerical results presented above we see that 
the profiles of the localized solutions of Eq. (\ref{npse})
both in the single-site (strong attraction, large values of $g$) 
and in the multi-site soliton cases but with a weak transverse 
modulation (small values of $\alpha$) are smooth. 
This observation suggests that one could achieve some analytical 
insight into the model with
attraction by using a VA with the Gaussian ansatz (a review of the VA can
be found in Ref. \cite{progress}),
\begin{equation}
\psi (\mathbf{r})={\frac{1}{\pi ^{3/4}\sigma \eta^{1/2}}}\exp {\left\{ -{%
\frac{(x^{2}+y^{2})}{2\sigma ^{2}}}\right\} }\exp {\left\{ -{\frac{z^{2}}{%
2\eta ^{2}}}\right\} }\;,  \label{ansatz}
\end{equation}%
where $\sigma $ and $\eta $ are variational parameters accounting for the
transverse and axial width of the BEC. Inserting this ansatz in Eqs. (\ref%
{efun}) and (\ref{mu}), with $V(z)=0$ (and $\gamma \equiv -g$), we obtain
the respective expressions for the energy-per-atom functional and chemical
potential,
\begin{equation}
E={\frac{1}{2\sigma ^{2}}}+{\frac{1}{4\eta ^{2}}}+{\frac{1}{2}}\left[
1-\alpha \exp {(-k^{2}\eta ^{2})}\right] \sigma ^{2}-{\frac{g}{2(2\pi )^{1/2}%
}}{\frac{1}{\sigma ^{2}\eta }}\;,  \label{E}
\end{equation}%
\begin{equation}
\mu =E-{\frac{g}{2\sqrt{2\pi }\sigma ^{2}\eta }.}  \label{muE}
\end{equation}%
Next, we minimize the energy, demanding $\partial E/\partial \sigma
=\partial E/\partial \eta =0$, which yields the variational equations,
\begin{equation}
\left[ 1-\alpha \exp {(-k^{2}\eta ^{2})}\right] \sigma ^{4}=1-{\frac{g}{%
(2\pi )^{1/2}\eta }}\;,  \label{var1}
\end{equation}%
\begin{equation}
2\alpha k^{2}\exp {(-k^{2}\eta ^{2})}\eta ^{4}={\frac{1}{\sigma ^{2}}}-{%
\frac{g\eta }{(2\pi )^{1/2}\sigma ^{4}}},  \label{var2}
\end{equation}%
which can be easily solved numerically \cite{method}. The solutions provide
for a minimum of the energy only if the curvature of the energy surface, $%
E(\eta ,\sigma )$, is positive, i.e.,
\begin{equation}
{\frac{\partial ^{2}E}{\partial \eta ^{2}}}{\frac{\partial ^{2}E}{\partial
\sigma ^{2}}}-\left( {\frac{\partial ^{2}E}{\partial \eta \partial \sigma }}%
\right) ^{2}>0\;.  \label{curvature}
\end{equation}

As concerns the dynamical stability of the solitons against small
perturbations, it may be, first of all, estimated by means of the VK
criterion \cite{VK}. According to it, a necessary stability condition is $%
d\mu /dg<0$ (in the present notation), if the soliton family is described by
dependence $\mu (g)$ (note that the VK criterion does not apply to gap
solitons in the model with the repulsive nonlinearity, therefore it was not
used in the previous section).

In Fig. 6, we plot axial length $\eta $ and transverse width $\sigma $
versus interaction strength $g$ for $k=1$ and four different fixed values of
modulation parameter $\alpha $. The figure shows that the soliton in the
self-attractive BEC exists up to a critical value of the nonlinearity
strength, $g_{c}$. At $g>g_{c}$, the 3D collapse of the nearly-1D soliton
occurs, which is a well-known result in the case of $\alpha =0$ \cite%
{PerezGarcia98,sala1,sala2}. Note that, for $\alpha =0$, the VA predicts $%
g_{c}=1.55$, which is somewhat higher than $g_{c}=1.34$ obtained from the
numerical solution of the full GPE in three dimensions \cite{gammal,sala2}.

\begin{figure}[tbp]
{\includegraphics[height=2.4in,clip]{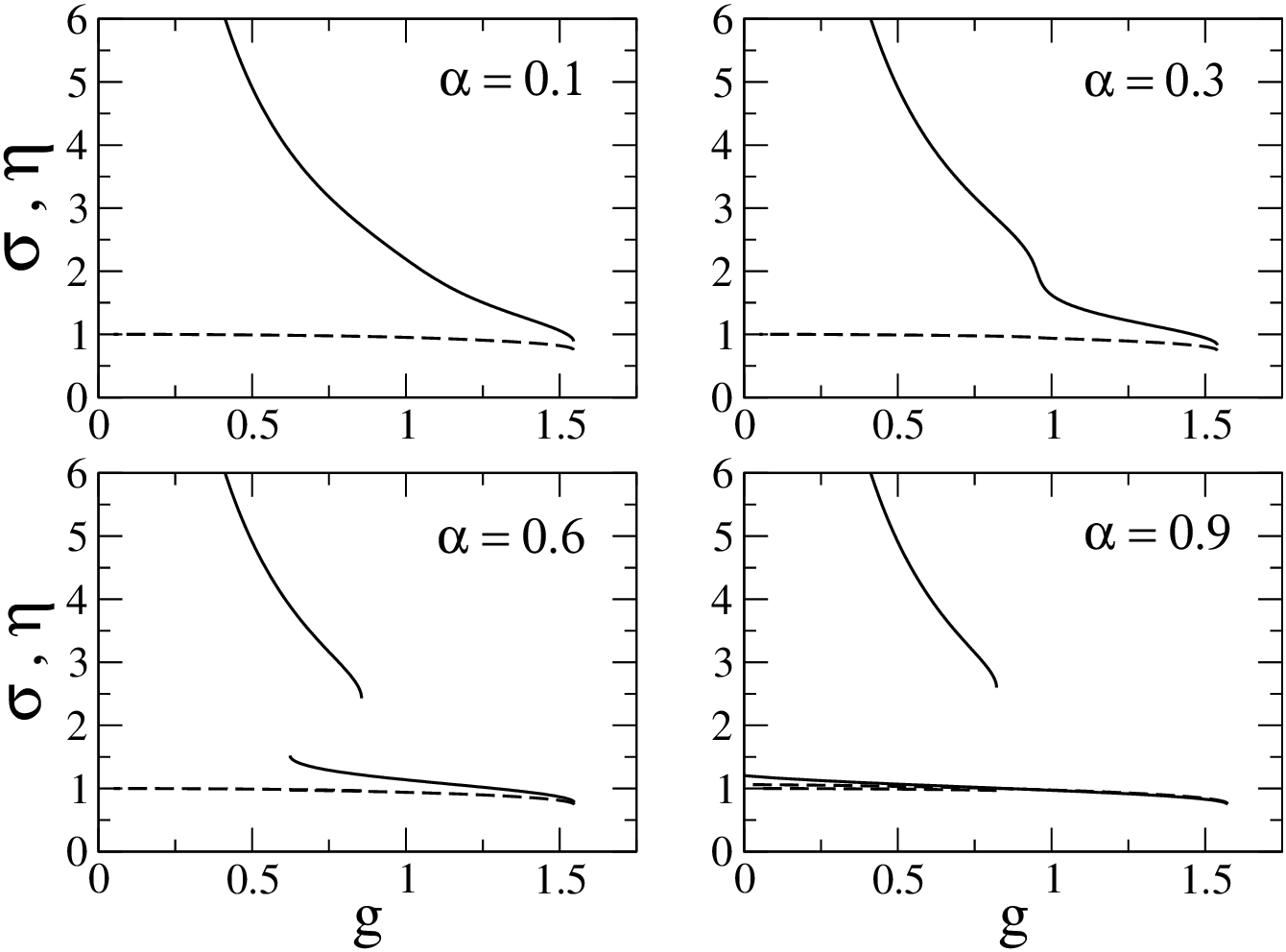}}
\caption{Transverse width $\protect\sigma $ (dashed lines) and axial length $%
\protect\eta $ (solid lines) of the quasi-1D bright soliton in the model
with attraction, versus the nonlinearity strength, $g\equiv -\protect\gamma %
=2|a_{s}|N/a_{\bot }$, as predicted by the variational approximation based
on ansatz (\protect\ref{ansatz}) and Eqs. (\protect\ref{var1}), (\protect\ref%
{var2}). On the right side, all curves terminate at the critical point, $%
g=g_{c}$, past which the collapse occurs. The dependences are displayed for $%
k=1$ and different values of the modulation depth $\protect\alpha $.}
\end{figure}

The axial length of the soliton, $\eta $, diverges as $g$ drops to zero,
while the transverse width $\sigma $ approaches $1$, actually becoming equal
to the above-mentioned harmonic-oscillator length, $a_{\bot }$. On the other
hand, as $g$ approaches $g_{c}$, both $\eta $ and $\sigma $ remain finite
and smaller than $1$.

New results are presented in Fig. 6 for $\alpha \neq 0$ (recall previous
works were only dealing with the case of $\alpha =0$). At small $\alpha $,
the figure shows only a slight distortion of the curves. A qualitative
change is observed at $\alpha >0.36$, when there appear two stable branches
for both $\sigma(g)$ and $\eta (g)$, the curves for $\eta (g)$ displaying a
clear gap. The lower branches of the $\eta (g)$ and $\sigma (g)$ dependences
exists only in a finite interval, which we denote as
\begin{equation}
g_{m}<g<g_{c},  \label{existence}
\end{equation}
while the upper branches extend up to $g=0$, 
in interval $0<g<g_{M}$, with $g_{M}<g_{c}$. 
Physically, the lower branches (with smaller values of axial
length $\eta$) correspond to single-site solutions, 
where the self-attractive BEC  is strongly
localized -- essentially, within a single cell of the modulation 
structure. The upper branches of $\eta (g)$ and $\sigma (g)$ 
correspond instead to the weakly localized solutions (with a
larger axial length), which occupy several cells (sites).

Analysis of expressions (\ref{E}) and (\ref{muE}) demonstrates that, in
interval (\ref{existence}), the multi-site and single-site solution may
assume the role of ground state (the one corresponding to the lowest
energy). However, this analysis also suggests that both families are \emph{%
dynamically stable}, as they always meet the VK criterion, $d\mu /dg<0$.
Direct numerical simulations (not shown in detail here) have confirmed this
conjecture.

For $\alpha >0.86$, the numerical solution of Eqs. (\ref{var1}) and (\ref%
{var2}) demonstrates that the lower border of existence interval (\ref%
{existence}) for the single-site soliton, $g_{m}$, vanishes, but this is an
artifact of the VA, which occurs in other contexts too \cite{Salerno}. It is
explained by the above-mentioned inadequacy of the Gaussian ansatz in the
limit of weak nonlinearity, i.e., for widely spread small-amplitude solitons
featuring a multi-peaked shape. In this situation, one may, in principle,
apply a more sophisticated ansatz, combining the Gaussian and periodic
functions, such as $\cos (2k_{L}z)$; however, the generalized ansatz results
in a cumbersome algebra \cite{Arik}, therefore we do not pursue such an
approach here.

In the upper panel of Fig. 7, we plot the density profile, $\rho (z)$, of
the soliton for different values of the self-attraction strength, $g$, and
fixed modulation parameters, $\alpha =0.9$ and $k=1$. As seen in the figure,
the increase of $g$ may strongly compress the soliton in the axial
direction, making the secondary maxima very small. In this case, the
condensate actually self-traps into a single-peak soliton, which occupies
only one cell of the modulation structure.

\begin{figure}[tbp]
{\includegraphics[height=2.7in,clip]{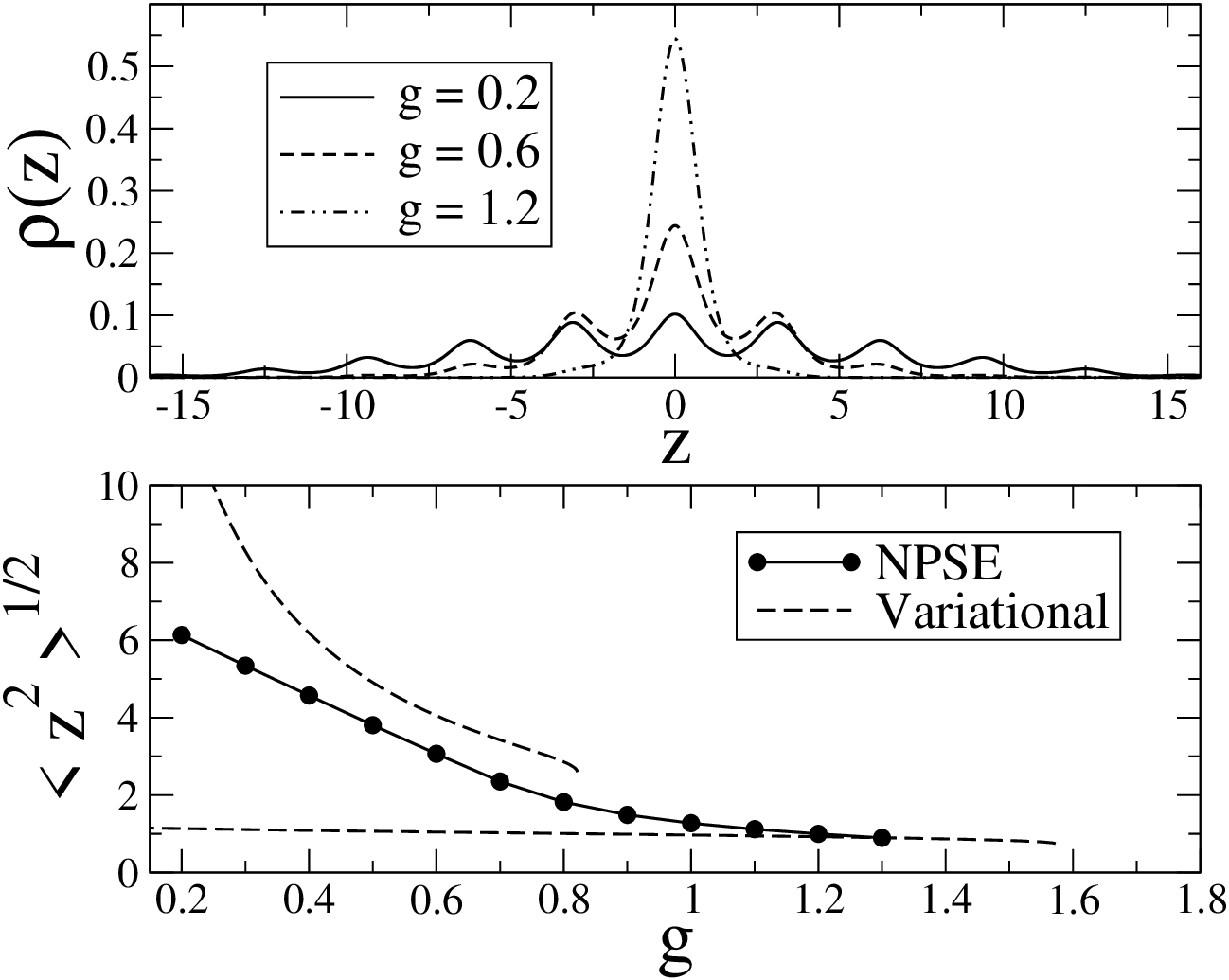}}
\caption{Upper panel: the axial density profile, $\protect\rho %
(z)=|f(z)|^{2} $, of the soliton in the attractive model, with $\protect%
\alpha =0.9$ and $k=1$, obtained for different values of the self-attraction
strength $g$ from the numerical solution of the NPSE, Eq. (\protect\ref{npse}%
). Lower panel: axial length of the soliton, $\left\langle
z^{2}\right\rangle ^{1/2}$, as a function of $g$, for $\protect\alpha =0.9$
and $k=1$. The lower panel includes results provided by the variational
approximation based on the Gaussian ansatz, see Eqs. (\protect\ref{var1})
and (\protect\ref{var2}), and those obtained from the numerical solution of
the NPSE.}
\end{figure}

Contrary to the numerical solution
of the NPSE which shows a smooth crossover between multi-site and single-site
solitons, the VA predicts a discontinuous transition. In the lower panel of Fig. 7, we plot the axial length, $\sqrt{%
\left\langle z^{2}\right\rangle }$, of the ground-state bright soliton as a
function of strength $g$, for $\alpha =0.9$ and $k=1$. 
The jump predicted by the variational calculation at $g\simeq 0.5$, is a consequence of the inadequacy of 
of ansatz (\ref{ansatz}), which assumes the simple Gaussian waveform
for the axial wave function, to describe  multi-peaked
states, as was also recently shown in the study of the quasi-1D model with
the self-attraction and axial optical lattice \cite{we}.

As it is well known, cold Bose atoms with attractive interactions 
collapse if the interaction strength exceeds a threshold value, 
$g_{c}$. Obviously, as $g$ increases towards $g_c$, 
the profile becomes narrower and narrower, and therefore close 
to collapse in the presence of a transverse modulation, its shape should 
correspond to a single-site soliton, where the VA is appropriate. 
It is thus interesting to predict the collapse threshold, $g_{c}$, 
as a function of parameters $\alpha$ and $k$ of the transverse 
modulation by using our gaussian variational ansatz. In Fig. 8, we 
display the dependence $g_{c}(k)$ predicted by the VA at five fixed values
of modulation depth $\alpha $. For given $\alpha $, the critical value $g_{c}
$ has its maximum at $k=0$, which is natural, as Eq. (\ref{U}) yields, in
this case, the smallest constant value of the transverse-trapping frequency.
Equation (\ref{U}) also helps to understand another feature observed in Fig.
8., \textit{viz}., that $g_{c}$ slowly diverges at $k=0$ as $\alpha $
approaches $1$ (for instance, $g_{c}=4.88$ at $\alpha =0.99$). Note also
that there exists a modulation wavenumber, $k_{c}$, at which $g_{c}$ attains
its \textit{minimum}, i.e., the collapse has the lowest threshold. Figure 8
shows that this minimum decreases with the increase of $\alpha $, which may
be understood too: as mentioned above, the strong potential tends to squeeze
the entire condensate into a single cell of the modulation structure, which
facilitates the onset of the collapse. On the other hand, at large values of
$k$, $g_{c}$ becomes asymptotically constant, as the interaction of the
condensate with the short-period modulation becomes exponentially weak,
hence it produces little effect on the collapse threshold.

\begin{figure}[tbp]
{\includegraphics[height=2.4in,clip]{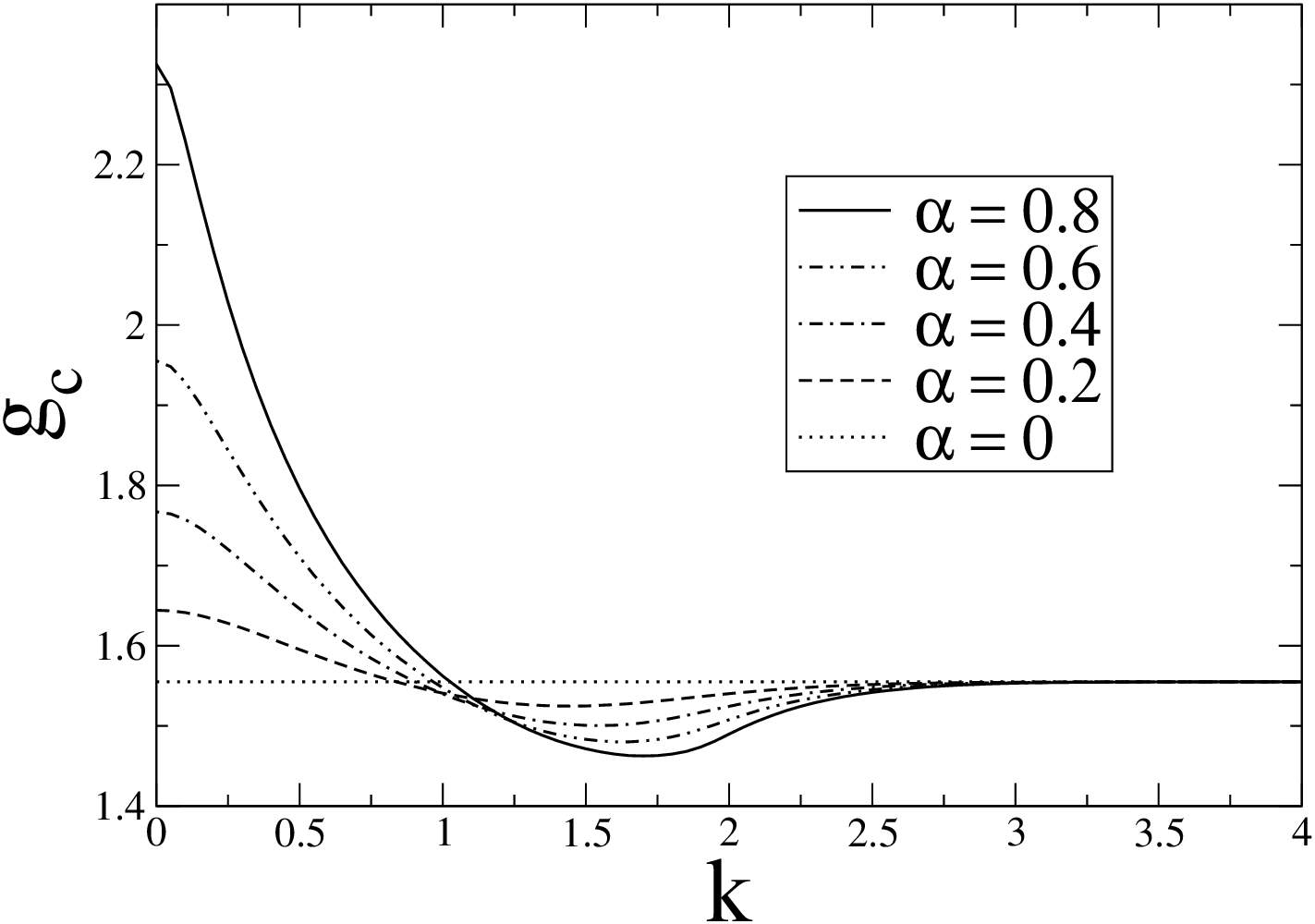}}
\caption{Critical strength $g_{c}$ for the collapse of the quasi-1D soliton
in the model with attraction versus modulation wave number $k$, as predicted
by the variational approximation. The dependences are displayed at fixed
values of modulation depth $\protect\alpha $.}
\end{figure}

In Table 1, we display the comparison of the VA-predicted critical value, $%
g_{c}$, to results following from the numerical solution of NPSE (\ref{npse}%
) (for $\alpha =0.9$). The variational and numerically found critical values
are seen to be in qualitative agreement. Note that, at $k=0$, the critical
value for the NPSE is $g_{c}={(4/3)}\left( 1-\alpha \right) ^{-1/4}$, which
is an obvious extension of $g_{c}=4/3$ found before at $\alpha =0$ \cite%
{sala1,sala2}.

{\vskip 0.3cm}

\begin{center}
\begin{tabular}{|c|c|c|c|c|}
\hline\hline
~~~$k$~~~ & ~~~$g_{c}$~~~~ & ~~~$g_{c}^{\mathrm{(var)}}$~~~~ & ~~~{\small $%
\sqrt{\left\langle z^{2}\right\rangle }$}~~~ & ~~~$\sigma (0)$~~~ \\ \hline
0.00 & 2.12 & 2.76 & 2.16 & 0.86 \\
0.10 & 1.99 & 2.52 & 1.20 & 0.72 \\
0.25 & 1.74 & 2.17 & 0.91 & 0.65 \\
0.50 & 1.55 & 1.85 & 0.75 & 0.56 \\
1.00 & 1.41 & 1.57 & 0.76 & 0.57 \\
1.50 & 1.34 & 1.47 & 1.19 & 0.79 \\
2.00 & 1.32 & 1.48 & 1.25 & 0.83 \\
2.50 & 1.45 & 1.54 & 0.99 & 0.75 \\
3.00 & 1.44 & 1.55 & 0.93 & 0.73 \\
3.50 & 1.42 & 1.55 & 0.92 & 0.73 \\ \hline\hline
\end{tabular}
\end{center}

{\small Table 1. Properties of the bright soliton in the model with
attraction in a proximity to the collapse, as found from the numerical
solution of Eq. (\ref{npse}) at $\alpha =0.9$ and different values of
modulation wavenumber $k$: $g_{c}$ is the critical value of the nonlinearity
coefficient at the collapse point, }$\sqrt{{\small <z^{2}>}}${\small \ is
the axial width of the soliton, and $\sigma (0)$ its transverse width (at $%
z=0$) at the same critical point. For comparison, also included are values
of $g_{c}$ predicted by the variational approximation, }$g_{c}^{\mathrm{(var)%
}}${\small .}

{\vskip0.3cm}

Table 1 demonstrates that, with the increase of $k$, the critical value, $%
g_{c}$, drops from the largest value, corresponding to $k=0$, to a minimum
at $k=2$, and then slightly increases with the further increase of $k$ (the
minimum at $k=2$ seems quite shallow from the side of $k>2$). This feature
and the related ones can be explained if one notices that $k=0$ corresponds
to a constant value of the modulation factor in Eqs. (\ref{npse}) and (\ref%
{sigma-npse}),
\begin{equation}
\beta _{k=0}\equiv \sqrt{1-\alpha \cos \left( 2kz\right) }|_{k=0}=\sqrt{%
1-\alpha },  \label{k=0}
\end{equation}%
and, on the other hand, for large $k$ (formally, for $k\rightarrow \infty $%
), the modulation factor should be replaced by its average,
\begin{equation}
\beta _{k=\infty }\equiv \left\langle \sqrt{1-\alpha \cos \left( 2kz\right) }%
\right\rangle =\frac{2}{\pi }\sqrt{1+\alpha }\, \mathbf{E}\left( \sqrt{\frac{%
2\alpha }{1+\alpha }}\right) ,  \label{k=infinity}
\end{equation}%
where $\mathbf{E}$ is the complete elliptic integral of the second kind.
Then, Eqs. (\ref{npse}) and (\ref{sigma-npse}) with the constant modulation
factor $\beta $ (and without the extra potential, $V=0$), take the form of%
\[
\left[ -{\frac{1}{2\beta }}{\frac{d^{2}}{dz^{2}}}+{\frac{1}{2}}\left( {\frac{%
1}{\sqrt{1-g|f(z)|^{2}}}}+\sqrt{1-g|f(z)|^{2}}\right) \right.
\]%
\begin{equation}
\left. -{\frac{{g}|f(z)|^{2}}{\sqrt{1-g|f(z)|^{2}}}}\right] f=\frac{\mu }{%
\beta }\,f  \label{beta}
\end{equation}%
(recall $g\equiv -\gamma $). Further, the coefficient $\beta$ can be
eliminated from Eq. (\ref{beta}) by means of an obvious rescaling [which
also takes into regard the condition that the normalization of the solution
must keep the form of Eq. (\ref{N1D})]:%
\[
\sqrt{\beta }z\equiv \tilde{z},~\mu /\beta \equiv \tilde{\mu},~\beta
^{-1/4}f\equiv \tilde{f},~\sqrt{\beta }g=\tilde{g}.
\]%
As a consequence, the critical values of $g$, together with the respective
length scale, which correspond to the different constant values of $\beta $,
are related as follows:%
\begin{equation}
\frac{\left( g_{c}\right) _{k=0}}{\left( g_{c}\right) _{k=\infty }}=\frac{%
\left( {\small \sqrt{\left\langle z^{2}\right\rangle }}\right) _{k=0}}{%
\left( {\small \sqrt{\left\langle z^{2}\right\rangle }}\right) _{k=\infty }}=%
\sqrt{\frac{\beta _{k=0}}{\beta _{k=\infty }}}.  \label{ratio}
\end{equation}%
For $\alpha =0.9$ (the value for which numerical data are collected in Table
1), Eqs. (\ref{k=0}) and (\ref{k=infinity}) yield $\beta _{0}\approx $ $%
\allowbreak 0.32$ and $\beta _{\infty }\approx 0.93$, hence the ratios
predicted by Eq. (\ref{ratio}) take the value $\approx 0.59$. On the other
hand, the numerical data from Table 1 yield $\left( g_{c}\right)
_{k=2}/\left( g_{c}\right) _{k=0}\approx \allowbreak 0.62$, and $\left(
{\small \sqrt{\left\langle z^{2}\right\rangle }}\right) _{k=2}{\LARGE /}%
\left( {\small \sqrt{\left\langle z^{2}\right\rangle }}\right) _{k=0}\approx
$ $\allowbreak 0.58$. Thus, approximating $g_{c}$ at $k=2$ by $\left(
g_{c}\right) _{k=\infty }$ (recall the change of $g_{c}$ at $k>2$ is
insignificant), one can explain (at least, in a crude approximation) effects
caused by the transition from long-scale to short-scale spatial modulation.

\subsection{Solitons in the first finite bandgap}

The above analysis was dealing with solitons (in the attractive model) whose
chemical potential belongs to the semi-infinite bandgap in the linear
spectrum of Eq. (\ref{npse}). On the other hand, it is known that the cubic
self-attractive nonlinearity may also give rise to solitons located in
higher-order (finite) bandgaps. Here, we report soliton solutions of the
latter type, found from the NPSE by means of the self-consistent numerical
method with periodic boundary conditions, which was outlined in the previous
section.

In Fig. 9 we plot the first $50$ eigenvalues $\mu _{j}$, as found from the
numerical solution of the stationary nonlinear NPSE, versus the nonlinearity
strength $g$. The first $32$ eigenstates form the first band, and the other $%
18$, which are well separated by a gap, cluster into the second band. Figure
9 shows that the lowest eigenvalue and the 33rd one split off from the first
and second bands and move down (till the onset of the collapse) with the
increase of $g$, thus giving rise to localized states in the semi-infinite
and first finite gaps, respectively. It is worthy to note that the second
eigenvalue, which originally belonged to the first band, also splits off
from it at larger values of $g$. We have verified that the corresponding
nonlinear eigenstate become localized, as one may expect. Qualitatively
similar findings were reported in Ref. \cite{efremidis}, which was dealing
with a numerical solution of the ordinary cubic GPE in one dimension, and,
more recently, in Ref. \cite{we} which was dealing with the self-attractive
BEC in the quasi-1D trap with an axial OL potential.

\begin{figure}[tbp]
{\includegraphics[height=2.4in,clip]{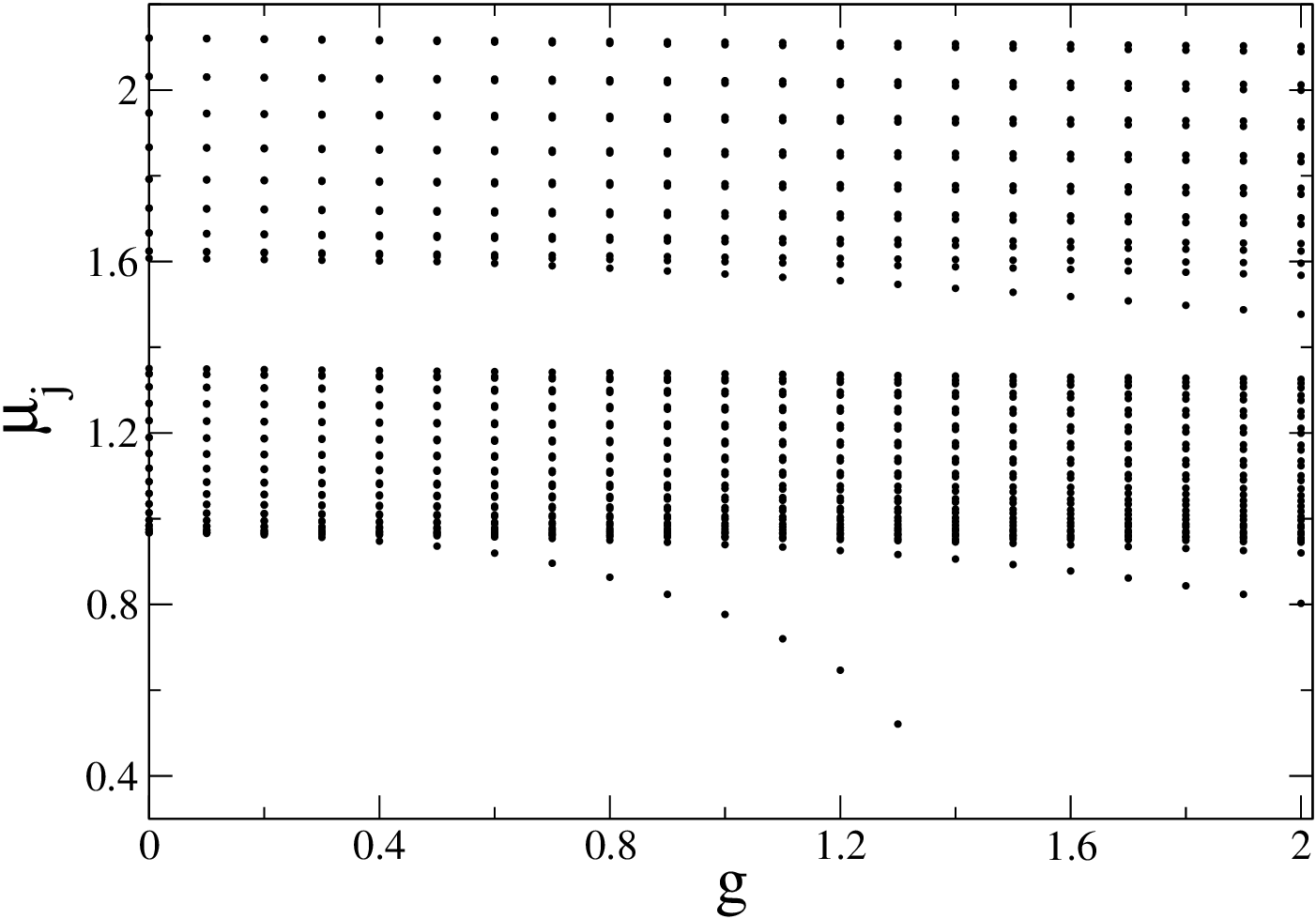}}
\caption{First 50 eigenvalues $\protect\mu _{j}$ found from the numerical
solution of NPSE, Eq. (\protect\ref{npse}), with $\protect\alpha =0.5$ and $%
k=1$, as functions of the nonlinearity strength in the model with
attraction, $g\equiv -\protect\gamma $.}
\end{figure}

The density profiles, $|f_{33}(z)|^{2}$, of the 33th state which develops
into a soliton belonging to the first finite bandgap, are displayed, for $%
\alpha =0.5$ and six different values of interaction strength $g$, in Fig.
10. The figure shows that this state is still fully delocalized (being
similar to a Bloch wave) at $g=0.4$, while at $g=0.6$ it becomes localized,
with the width much smaller than the length of the periodic box, featuring
many local maxima and minima (zeros). With further increase of $g$, the gap
soliton keeps compressing itself. Comparing Figs. 4 and 10, we conclude that
the ground-state solitons, which reside in the semi-infinite gap, are
drastically different from the \textit{gap solitons}, i.e., ones found in
the first finite bandgap. First, for the same parameters, the ground-state
solitons are localized much stronger, and their local density minima are not
zeros, unlike those of the gap solitons. In addition, it is noteworthy that
local density maxima of the ground-state solitons correspond to minima of
the effective periodic potential, while the local maxima of the gap soliton
correspond to \emph{maxima} of the periodic potential.

\begin{figure}[tbp]
{\includegraphics[height=2.3in,clip]{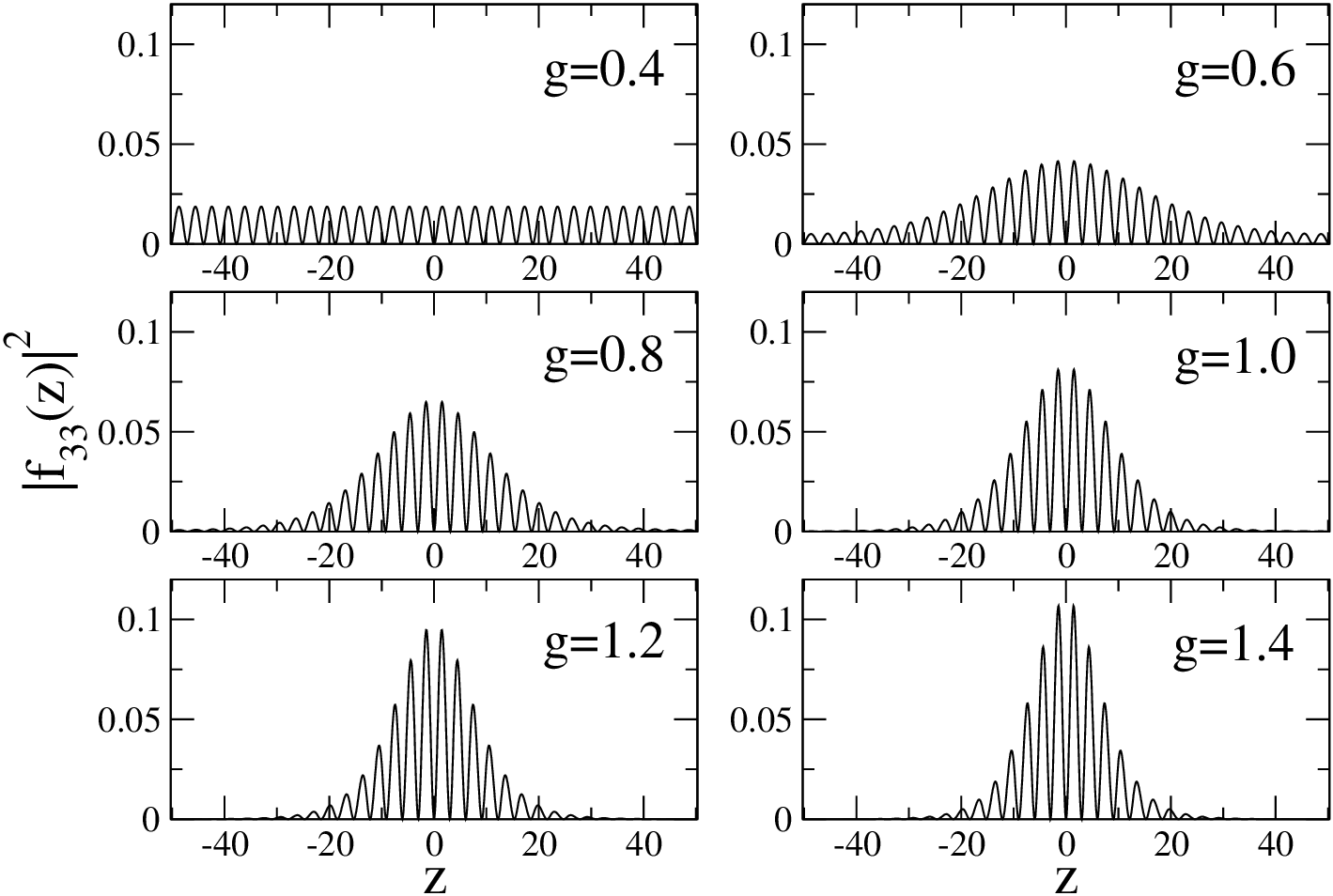}}
\caption{Axial density profiles, $|f_{33}(z)|^{2}$, of the 33th nonlinear
eigenfunction of the NPSE (which develops from a Bloch-like delocalized
state into a gap soliton) for $\protect\alpha =0.5$, $k=1$ and six values of
$g$.}
\end{figure}

Although the gap solitons cannot play the role of ground states, they, as
well as the ground-state solitons, are stable in direct simulations of the
time-dependent variant of the NPSE equation. Therefore, the gap-soliton
states are relevant to the experiment.

\section{Conclusions}

In this work, we have derived the effective one-dimensional NPSE
(nonpolynomial Schr\"{o}dinger equation) for a cigar-shaped trap
whose transverse confining frequency is periodically modulated
along the axial direction, thus inducing an effective periodic
axial potential. Besides the usual quasi-1D linear geometry, the
model may also be relevant as a means of creating an effective
periodic potential in toroidal traps. In both cases of the
repulsive and attractive nonlinearity, delocalized states and
solitons were found, by means of numerical methods (which were
applied to both the effective NPSE and the underlying 3D
Gross-Pitaevskii equation) and VA (variational approximation; this
method was applied to ground-state solitons in the model with
attraction, and to the prediction of the collapse threshold in
this case). It was found that the numerical solution to the NPSE
is always extremely close to the full 3D solutions. The VA yields
quite reasonable results too, except for the description of the
crossover from single-site to multi-site solitons: numerical
results reveal that the crossover is smooth and does not include a
jump, contrary to the prediction of the VA. This shortcoming of
the VA is explained by the fact that the simple Gaussian ansatz,
on which the approximation is based, cannot adequately grasp the
transition that alters the shape of the soliton, giving it the
multi-peaked structure. The transition from delocalized states to
gap solitons was studied in detail (by means of numerical methods)
in the first finite bandgap, for both cases of the repulsive and
attractive nonlinearities.

The above results, presented in terms of the dimensionless equations, can be
easily translated into physical units. For instance, by considering an
attractive Bose-Einstein condensate made of $^{7}$Li atoms, with scattering
length $a_{s}=-1.45$ nm, and choosing the transverse confining frequency as $%
\omega _{\bot }=2\pi \times 100$ Hz, we have a typical value of the
modulation period, $\lambda =\pi a_{\bot }/k=12$ $\mu $m, for $k=1$ (recall
it was a typical value for which the results were reported above). In this
case the critical number of atoms at the collapse threshold is $%
N_{c}=g_{c}a_{\bot }/(2|a_{s}|)\approx 1850$.

The results may be quite useful to design new experiments in
toroidal traps, where an azimuthal periodic potential can be
induced, as explained above, by the spatially-modulated transverse
confinement.

The work of B.A.M. was partially supported by the Israel Science Foundation
through Excellence-Center Grant No. 8006/03, and by German-Israel Foundation
through Grant No. 149/2006. L.S. acknowledges GNFM-INdAM for partial
support. L.S. and L.R. thank Prof. Alberto Parola for numerous discussions
and suggestions.


\begin{thebibliography}{99}
\bibitem{method} Equations (\ref{var1}) and (\ref{var2}) are solved by
choosing $\eta $ as an independent parameter. Then, defining $A=\alpha \exp {%
(-k^{2}\eta ^{2})}$ and $B=A(1+2k^{2}\eta ^{2})-1$, one finds, after some
algebraic manipulations, $\sigma ^{2}=\left( 1-\sqrt{1-4\eta ^{4}B}\right)
/\left( 2\eta ^{2}B\right) $, and $g=\eta (1-(1-A)\sigma ^{4})/\sqrt{2\pi }$.
\end{thebibliography}
\end{document}